\documentstyle[epsfig,preprint,aps]{revtex}
\begin{document}
\def\bra#1{{\langle #1{\left| \right.}}}
\def\ket#1{{{\left.\right|} #1\rangle}}
\def\bfgreek#1{ \mbox{\boldmath$#1$}}

\title{Pseudovector versus pseudoscalar coupling \\in kaon photoproduction --- revisited}
\author{S.S. Hsiao$^1$, D.H. Lu$^2$ and Shin Nan Yang$^2$}
\address{$^1$Department of physics, Soo Chow University,
	Taipei, Taiwan 11102}
\address{$^2$Department of Physics,
	National Taiwan University, Taipei, Taiwan 10617}
\maketitle

\begin{abstract}
The question of pseudovector versus pseudoscalar coupling schemes for the
kaon-hyperon-nucleon interaction is re-examined for the reaction
$\gamma p\to K^+ \Lambda$ in several isobaric models. These models typically  
include Born terms, $K^*$- and $K_1$-exchange in the t-channel, and a few
different combinations of spin-1/2 baryon resonances 
in the $s$- and $u$-channels.
The coupling constants are obtained by fitting to a large data set. 
We find that both pseudoscalar and pseudovector couplings can
allow for a satisfactory description of the present database.
The resulting coupling constants, $g_{K\Lambda N}$ and $g_{K\Sigma N}$,
 in  the pseudovector coupling scheme are smaller
than those predicted using flavor SU(3) symmetry, 
but consistent with the values obtained in a QCD sum rule calculation.
\end{abstract}
\noindent
PACS numbers: 13.60.Le, 25.20.Lj
\newpage
Kaon electromagnetic production has been studied for 
more than three decades.
However, the progress has not been as swift as in the
case of pion production.
It is due mostly to the lack of precise experimental data. This is
changing as abundant data are coming out from various high energy,
high duty cycle electron accelerators like TJNAF, ELSA and ESRF.

 On the theoretical side, most of the calculations have employed the isobar model
approach \cite{WJC,SL96,Bennhold99,Feuster99}. 
Such an approach includes a
limited number of low-lying $s$-, $t$-, and $u$-channel resonances,
together with the Born terms, in a fit to  data. These
phenomenological analyses have been hampered by the fact that many
resonances can, in principle, contribute due to the large energy needed to
produce a kaon. They differ from each other mostly in the particular set of
resonances considered. Despite many persistent efforts to reproduce
 data~\cite{WJC,SL96,Bennhold99,Feuster99}, 
serious problems remain in the description
of  kaon production.
For example, the coupling constant $g_{K\Lambda N}/\sqrt{4\pi}$
obtained from the fits by
Adelseck-Saghai (AS)~\cite{AS}, Williams, Ji and Cotanch (WJC)~\cite{WJC} and
Mart, Bennhold and Hyde-Wright (MBH)~\cite{MBH95} are $-4.17\pm0.75, -2.38$,
and $0.51$, respectively, 
as compared with the SU(3) value of $-3.7\pm0.7$~\cite{AS}.

Most theoretical analyses performed so far have employed pseudoscalar 
coupling (PS)
for the kaon-hyperon-nucleon vertex. This is because~\cite{Bennhold89}
the use of pseudovector coupling would lead to a further suppression
of the leading Born couplings in the fit to data . Another reason is that
the value of the coupling constant $g_{K\Lambda N}/\sqrt{4\pi}$ obtained
from the fit within the pseudovector coupling scheme is, 
in general, considerably
smaller than the SU(3) value. However,	under flavor
SU(3) symmetry,  kaon is a member of the pseudoscalar
meson octet, as well as the pion and eta meson.
Thus, it is natural to expect that the kaon-hyperon-nucleon $(KYN)$ vertex
takes the same form as $\pi NN$.
In the $\pi N$ system, it is well-established that the
pseudovector (PV) coupling scheme has an advantage
over PS coupling as it respects current algebra and
incorporates  low energy theorems.
Furthermore, with SU(3) symmetry breaking effects taken into account,
a recent QCD sum rule calculation~\cite{Choe96} gave
$g_{K\Lambda N}/\sqrt{4\pi}=-1.96$, which is only about half of the SU(3)
value. We remind the readers that the result of
$g_{K\Lambda N}/\sqrt{4\pi}= -4.17\pm0.75$ obtained by
 AS \cite{AS}, which appears to agree well with the SU(3) value, was actually
 more of a constraint imposed in the fitting. 
They found many other possibilites within the PS coupling scheme which 
could give a comparable reduced $\chi^2$. It is 
clear then that the issue is far from being settled. Bennhold and Wright (BW)
\cite{Bennhold87} investigated this question
of PV versus PS coupling for the KYN vertex
in kaon photoproduction more than ten years ago. They concluded that the data
did not prefer
one coupling over the  other. However, only Born terms were included
in the model considered  by BW and the fitted data
were those available before 1984 which were rather limited.  Accordingly, 
we want to re-address this 
important question for kaon photoproduction   within  more extended 
models and
the larger database which is  currently available,
as recently called for by Bennhold et al.~\cite{Bennhold97}.

The extended models we considered  are similar to
those employed by AS~\cite{AS} and WJC~\cite{WJC}. 
They consist of Born terms, $K^*(892)$ and $K_1(1270)$ exchanges in 
the $t-$channel, and a number of spin-1/2 baryon resonances in the $s-$ 
and $u-$channels. The kaon-baryon-baryon
$(KBB')$ interaction, where $B$ and $B'$ can be $N, N^*, Y$ and $Y^*$,
 in either coupling scheme is given as follows:
\begin{eqnarray}
\protect{\cal{L}}^{PS}_{KBB'} &=& 
-g_{KB'B} \overline{\psi}_{B'}\Gamma_\pm\psi_B\,\phi_K, \\
\protect{\cal{L}}^{PV}_{KBB'} &=& 
{f_{KB'B}\over m_K}
\overline{\psi}_{B'}\gamma_\mu\Gamma_\pm \psi_B\,
\partial^\mu\phi_K, 
\end{eqnarray}
where $\Gamma_+ = i\gamma_5$ and $\Gamma_- = 1$, 
depending on whether $B'$ and $B$
have the same or opposite parity.
As in the $\pi N$ interaction, the ``equivalent'' coupling constant
for the  $KBB'$ in PV coupling is related to that in PS coupling
through the relation
\begin{equation}
{g_{KBB'}\over (m_B + m_{B'})} = {f_{KBB'}\over m_K}.
\end{equation}
In pseudoscalar coupling,
the Born terms are those given in Figs.~1(a)-1(c), while the additional
``seagull'' diagram  of Fig.~1(d) is needed in pseudovector coupling in order
to maintain gauge invariance. The couplings of vector mesons $K^*$ and $K_1$
with baryons are taken to be a sum of vector and tensor parts, as given in
Ref.~\cite{ABW85}. With the standard form for the electromagnetic vertices
$\gamma BB'$ and $\gamma MM'$, where $M(M')=K, K^*$ and $K_1$~\cite{ABW85},
it is straightforward to derive the resulting kaon photoproduction amplitude.
Explicit expressions for various amplitudes within the PS coupling scheme 
can be found
in Ref.~\cite{ABW85}. In our present calculation, the amplitudes in both PS
and PV coupling schemes are evaluated by a computer program which carries
out the Dirac algebra in helicity basis.

The first  model (I) we consider is that employed by AS~\cite{AS}
which includes the Roper resonance $N(1440)$ (N1) and $\Lambda(1670)$ (L3)
 in the $s-$ and
$u-$channels, respectively. We follow the notation, e.g., N1 and L3 used
above, of Ref. \cite{SL96} to denote various baryon resonances. In the second
model (II), one more resonance $\Lambda(1405)$ (L1) in the $u-$channel
is added to AS model. As can be seen in Table I,
where the  meson and baryon resonances included in each model are listed,
the third model (III) we study
 contains four more resoances, i.e., $N(1650)$ (N4), $N(1710)$ (N6),
$\Lambda(1750)$ (L5) and $\Sigma(1660)$ (S1), than model II.

The fitted data set used in the BW's study of PS {\em vs} 
PV coupling~\cite{Bennhold87}
consists of 131 data points for the photon laboratory energy $E_\gamma$
in the range of $930 - 1400$ MeV, all for the reaction of
$p(\gamma,K^+)\Lambda$. Of these 131 data points, 108 of them are
differential cross sections while the rest are polarization data.
In the present study,  242 data points (cross sections and polarization)
from the $\gamma p\rightarrow
K^+\Lambda$ reaction are used in the fitting procedure, as used in the 
calculation of Ref.~\cite{SL96}.

The resulting parameters obtained in the least-squared fit to the data
and the chi-square per degree of freedom within both PS and PV coupling
schemes for the three models described above are listed in Table I.
In several cases certain combinations of strong and electromagnetic 
couplings, e.g., $g_{K\Lambda N^*}\kappa(NN^*)$, 
where  $\kappa(NN^*)$ is the transition
magnetice moment of $NN^*$, always arise together. Therefore,
only their product like $G_{N^*}=g_{K\Lambda N^*}\kappa(NN^*) $ can be
determined in current study as given in Table I.
We first repeat the calculation of AS \cite{AS}, which used PS coupling and
 fitted to only 117 differential cross section data for the reaction
 $p(\gamma,K^+)\Lambda$ available at that time. We find a set of
 coupling contants which differ slightly from those of their Model A but
lead to a smaller $\chi^2/N = 1.21$ as shown in the first column of Table I.
We then employ the same model, i.e., including the Born terms and keeping
only N1 and L3 resonances but refit to a larger database of 242 data
points from the $p(\gamma,K^+)\Lambda$ reaction. The refitted 
 coupling constants are listed in the column denoted by PS-I.
The resulting
$\chi^2/N$ increases to $1.56$ since the number of data points considered is 
considerably larger. Many of the coupling constants  obtained differ 
substantially from the AS values , e.g.,  $g_{K\Lambda N}/\sqrt{4\pi}$ changes 
from $-4.11$ to $-1.55$.
Clearly the selection of database is very important in determining
the coupling constants.
As demonstrated in Ref.~\cite{Workman91}, we also find that
the coupling constant $g_{K\Sigma N}$
can not be determined by the data for the reaction
$\gamma p\rightarrow K^+\Lambda$ alone 
(even the sign of $g_{K\Sigma N}$ may  change).
The column labeled by PV-I gives the  results within the same Model I, 
but with the PV coupling scheme for
the $KBB'$ vertices.
It gives an almost identical $\chi^2/N$ as in PS-I, but
the resulting fundamental coupling constants,
$g_{K\Lambda N}/\sqrt{4\pi}$ and $g_{K\Sigma N}/\sqrt{4\pi}$ 
decreases by about $20\%$ as compared to PS-I value.

PS-II and PV-II columns give the results obtained with model II which
contains an additional hyperon resonance, $\Lambda(1405)$ (L1) as compared
to model I.
The addition of L1 strongly affects other coupling constants,
in particular,  $G_{K\Sigma N}$ and $G_{L3}(1670)$.
As in  model-I, the coupling constant $g_{K\Lambda N}/\sqrt{4\pi}$
in the PV scheme is smaller than that in the PS scheme.
We  have tried a number of combinations of baryon resonances
in our fitting process. We find that a reasonable $\chi^2/N$
can be achieved by several different models due to  the 
quality of the present database.
A typical result is presented  in 
the last two columns, PS-III and PV-III, of the table.
The $\chi^2/N$'s obtained with model III become smaller 
because four more resonances are
included.

In Fig.~\ref{fig_dcs} we show the differential cross sections for the above
three models as a function of the photon energy (Left)
and the scattering angle (Right).
Note that the points with open square are the latest data
from SAPHIR collaboration~\cite{SAPHIR98} and
are not included in our fitting procedure.
At lower energies,
both PS and PV schemes can provide reasonable descriptions, in other words,
the data do not distinguish PS and PV couplings in this region.
As the photon energy increases, the theoretical predictions
in the PS and PV schemes differ considerably.

In Fig.~\ref{fig_pol} we show recoil polarization of the $\Lambda$ with
respect to the photon energy (Left) and the scattering angle (Right).
Due to scarcity of data and large error bars, this quantity
gives a small contribution to $\chi^2$.
As in Fig.~\ref{fig_dcs} the deviations start mainly after 
$E_\gamma = 1.3$ GeV.
We would like to point out that the present simple
model is not able to reproduce the node structure in the angular
distribution of the $P_\Lambda$, as indicated by the recent data
from SAPHIR~\cite{SAPHIR98}.
Since $P_\Lambda$ is due to the interference  between helicity
amplitudes, resonances and final state interactions
play significant role. A quantitative fit to this observable
is possible only with refined models including
form factors and final state interactions.
For completeness, we last present the total cross section 
in Fig.~\ref{fig_tot}.
For photon energies below 1.5 GeV, both schemes work quite 
well. To reproduce
the higher energy data, it is essential to have hadronic
form factors at all interaction vertices \cite{Bennhold99,Lu95}.

Generally, the models with PS coupling give diversified results
for the fundamental coupling constants
(differing by up to a factor of 3).
This implies that the Born terms are not stable with respect to the addition
 of higher resonances in the PS scheme.
In contrast, the fitted results with
PV coupling are quite stable toward such additions,
and the leading coupling constants in PV schemes
are close to each other.
Note that the role of $N (1650)$ emphasized
by other group~\cite{Bennhold99,WJC,SL96,Workman91}
is not as explicit in our work.
The ability to reach a small $\chi^2$ in most of our cases indicates
that the neglect of higher spin resonances (spin-3/2 and higher)
is justified in the energy region in which we are interested.

In summary, we have tested the PS and PV schemes for the
kaon-baryon interaction in the  $\gamma + p \to K^+  + \Lambda$ reaction.
Our results show that the PV coupling scheme
for the kaon-hyperon-nucleon  can not be ruled out by the
present database.
Both schemes can provide reasonable
descriptions of the data for the differential cross section
below $E_\gamma = 1.5$ GeV.
The resulting coupling constants in the PV scheme are 
somewhat smaller
than those from the SU(3) limit,
but are consistent with values obtained from a
 QCD sum rule calculation~\cite{Choe96}.
To resolve this question, precise data, in particular 
$\Lambda$ polarization at backward angles will be helpful, 
together with a refined theoretical model with a proper treatment of
hadron size and final state interactions.

Another  possibility of examining the coupling scheme
is the study of kaon photoproduction from nuclear matter.
In this case, whether the leading order term proceeds through the
contact interaction, which only appears in the PV scheme,
or not could help to distinguish these two schemes.
Any contribution due to the PS coupling must rely on
the propagation of the nucleon or the hyperon inside 
the nuclear medium,
which shall manifest itself in the cross sections.

The authors thank B. Saghai for providing them with
data set used in this study and useful discussions.
This work is supported in part by the National Science Council
of ROC under grant No. NSC-89-2112-M002-038.

\newpage
\begin{table}[htbp]
\begin{center}
\caption{Exchanged particles and associated coupling constants.
From the QCD sum rule approach, the leading coupling constants,
$g_{K\Lambda N}/\sqrt{4\pi}$ and $g_{K\Sigma N}/\sqrt{4\pi}$,
are (-2.76, 0.44) for the SU(3) symmetric case
and become (-1.96,0.33) otherwise~\protect\cite{Choe96}.
Note that
$G_{N^*}\equiv g_{K\Lambda N^*} \kappa (N^*N)/\sqrt{4\pi}$ and
$G_{Y^*}\equiv g_{KY^*N}\kappa (Y^*\Lambda) /\sqrt{4\pi}$.}
\label{table1}
\vspace{1.0cm}
\begin{tabular}{c|c|c|cc|cc|cc}
particle&coupling & AS & PS-I&PV-I & PS-II&PV-II & PS-III&PV-III\\
\hline
$\Lambda$&$g_{K\Lambda N}/\sqrt{4\pi}$& -4.11 & -1.55 & -1.24 & -1.98 & -1.65 & -2.41 & -1.44\\
$\Sigma$ & $g_{K\Sigma N}/\sqrt{4\pi}$&  1.10 & 0.71 &	1.04 & -0.50 & 0.36 &	0.47 &	0.23\\
$K^*$(892)& $G_V/4\pi$	  & -0.44 & -0.13 & -0.11 & -0.14 & -0.21 & -0.17 & -0.17\\
	  & $G_T/4\pi$	  &  0.18 &  0.24 &  0.37 &  0.23 &  0.14 &  0.09 &  0.17\\
$K1$(1280) & $G_{V1}/4\pi$ & -0.10 & -0.17 & -0.22 & -0.17 & -0.07 & -0.18 &  -0.12\\
	   & $G_{T1}/4\pi$ & -1.13 & -0.13 & -0.07 & -0.29 & -0.30 & -0.36 & -0.23\\
\hline
$N(1440)$ & $G_{N1}$ & -1.43 & -1.25 & -1.11 &  -0.97 & -1.20 & -1.29 & -1.10\\
$N(1650)$ & $G_{N4}$ &       &       &       &       &       & -0.05 &  0.03\\
$N(1710)$ & $G_{N6}$ &       &       &       &       &       &  0.02 & 0.01\\
$\Lambda (1405)$& $G_{L1}$&	 &	 &	 &-0.06 & -0.78 & -0.08 & -0.51\\
$\Lambda (1670)$& $G_{L3}$& -3.09 & -0.09 & -1.38 &  -0.32 & -4.81 & -0.46 & -4.43\\
$\Lambda (1750)$& $G_{L5}$&	 &	 &	 &	 &	 & -1.81 &  0.25\\
$\Sigma (1660) $& $G_{S1}$&	&	&	&	&	&-0.42 & -0.45\\
$\chi^2/N$  &   & 1.21  & 1.56  &  1.57 &  1.56 &  1.46 & 1.38 &  1.38\\
\end{tabular}
\end{center}
\end{table}

\newpage
\begin{figure}[hbt]
\begin{center}
\epsfig{file=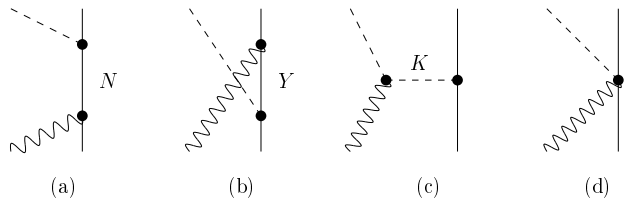,height=20cm}
\vspace{-14cm}
\caption{The Born term diagrams for $\gamma p\rightarrow K^+\Lambda$.}
\end{center}
\end{figure}

\begin{figure}[hbt]
\begin{center}
\epsfig{file=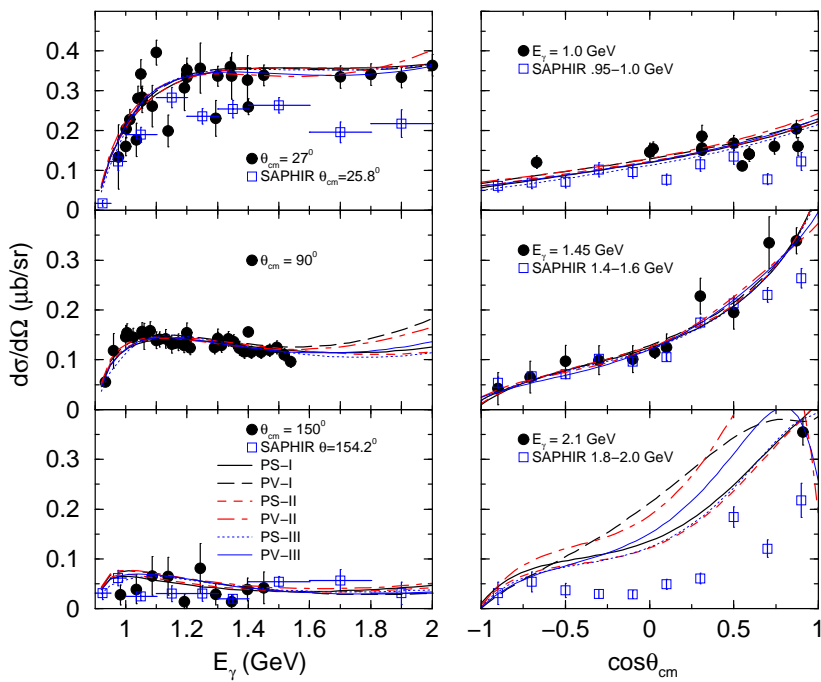,height=9cm}
\caption{Differential cross sections for the
$\gamma p\rightarrow K^+\Lambda$. The left side showes the energy
dependence at $\theta_{\rm cm}=27^0$, $90^0$ and $150^0$;
the right side gives angular distribution at $E_\gamma= 1.0$,
1.45 and 2.1 GeV, respectively.
The curves are model calculations for PS-I (solid),
PV-I(long-dashed), PS-II (dashed), PV-II (dot-dashed),
PS-III (dotted), and PV-III (thin-solid), respectively.
The data with filled circles are the same as in Ref.~\protect\cite{SL96}.
The data points with open squares are from SAPHIR
collaboration~\protect\cite{SAPHIR98}. }
\label{fig_dcs}
\end{center}
\end{figure}
\newpage
\begin{figure}[hbt]
\begin{center}
\epsfig{file=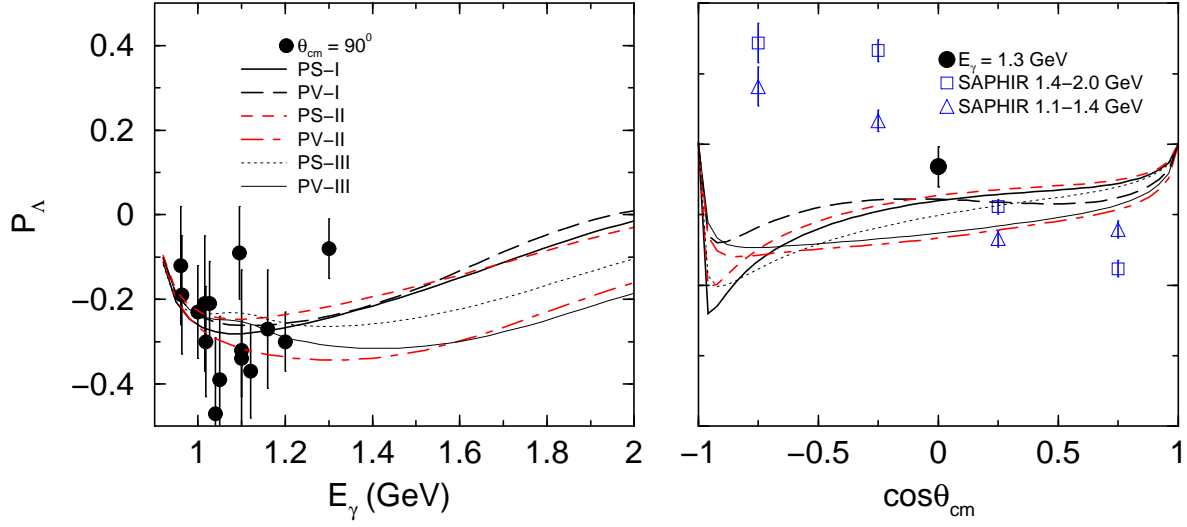,height=7cm}
\vspace{0.5cm}
\caption{The $\Lambda$ polarizations as a function of
the photon energy at $\theta_{\rm cm}=90^0$ (Left)
and a function of cos\,$\theta_{\rm cm}$
at $E_\gamma=1.45$ GeV (Right).
The legends for the curves and data are the same as in Fig.1. }
\label{fig_pol}
\end{center}
\end{figure}
\begin{figure}[hbt]
\begin{center}
\epsfig{file=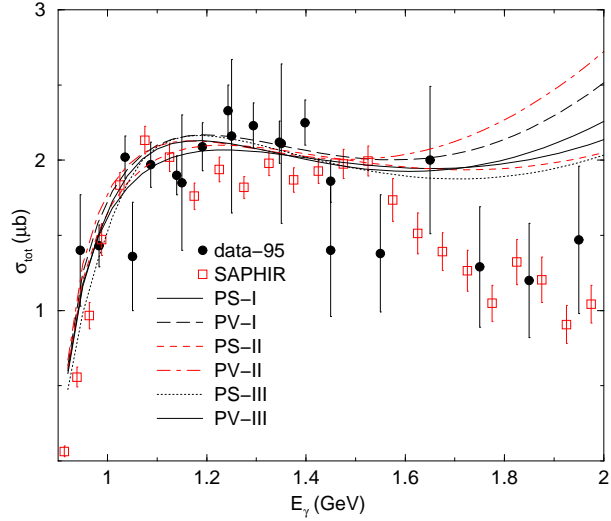,height=7cm}
\vspace{0.5cm}
\caption{Total cross sections as a function of the photon energy.
The legends for the curves and data are the same as in Fig.1. }
\label{fig_tot}
\end{center}
\end{figure}

\end{document}